**Kinetic fluid moments closure for a magnetized plasma with collisions**       12.07.18

R.E. Waltz, F.D. Halpern, Zhao Deng[1], and J. Candy

General Atomic, San Diego, CA 92121

*Abstract*

A novel method aimed at a kinetic moments closure for a magnetized plasma with arbitrary collisionality is proposed. The intended first application is to a tokamak edge and scrape-off-layer plasma. The velocity distribution function for each species is expanded in 8 Gaussian Radial Basis Functions (GRBFs) which are essentially shifted Maxwellians at eight representative 3D-velocity points of drift. The vector of 8 fluid moments (for particle density, 3 particle fluxes, total energy density, and 3 energy fluxes) has an 8x8 analytic linear matrix relation to the vector of 8 GRBF density weights in 3D-real space. The 8 fluid moments with sources for each species are advanced in time while the 8 GRBF weighs are determined from the 8x8 inverse matrix. The two closure moments (for the stress tensor and the energy weighted stress tensor) are linearly determined from the GRBF weights. Most importantly the velocity moments of the nonlinear Coulomb Fokker-Planck collision operator [Rosenbluth et al, Phys. Rev 107, 1957] are evaluated from the GRBF weights. Generalization from 8 to 12, 16, 20 .., in an energy weighted moment hierarchy is straightforward. The electric field follows from a generalized vorticity (quasi-neutral current continuity) equation. A *strong* drift ordering approximation can be applied to eliminate any spuriously unstable high frequency cyclotron motions. A novel *weak* drift ordering two-time step scheme avoids the vorticity equation by following ion cyclotron motion in time to get the electric field with ion gyro-averaging. Inclusion of low-beta magnetic perturbations is straightforward.

[1] Current address: Chinese Academy of Engineering Physics, Mianyang, China

Email: waltz@fusion.gat.com

## I. INTRODUCTION

It is well known that the kinetic (6D) equations for a plasma in a magnetic field with collisions can be represented by an infinite number of velocity or fluid moments. The symmetric moment hierarchy in a conservative time advance form is given in our principal reference text by Hazeline and Meiss[1] (see[Ref. 1, p.214]. The time advance of the first few (or finite number) of the lowest moments depends on the higher neglected moments which must be expressed in terms of the lower advanced moments: the so called *fluid moment closure* problem. The famous Braginskii[2] closure for a collisional magnetized ion and electron plasma truncates at 5 moments per species (particle density, 3 particle (or momentum) flux, and total energy density). The Braginskii closure for the collision moments and dependent transport fluxes is based on a short mean free path approximation questionable for the intended application of the novel closure method proposed here: numerical simulation of the turbulent tokamak edge and "scrap-off-layer". Ref. [3] extends and reviews previous work following the Braginskii closure while detailing the drift ordering, quasi-neutrality, and low-beta approximations commonly used in such codes (e.g. BOUT++[4], GBS[5]) This reference serves as a point of comparison and contrast with the "kinetic" fluid moments closure and methods proposed here. Going beyond the limits of validity on the Braginskii closure, the kinetic closure proposed here allows for arbitrary collisionality, flow strength, as well as strong deviation from drifted Maxwellian velocity distributions. However as in Ref. [3] and current collisional two-fluid tokamak edge simulation codes, the kinetic closure as applied here to the tokamak edge is first formulated with a *strong* drift approximation. This means that only low-frequency drift motions much less than ion cyclotron frequency and long perpendicular field length scales much greater than the ion gyroradius can be followed. A novel method with a *weaker* drift approximation is proposed which may be able to treat ion gyro-averaging at shorter perpendicular length scales by implicitly following the ion cyclotron motion in time. Cross field transport at electron gyroradius scales must be added with sub-grid scale diffusion models.

In this work, the kinetic closure is first illustrated for an 8 full moment system (particle density, 3 particle fluxes, total energy density, and 3 total energy fluxes) per species to be explicitly time advanced in conservative form with external sources (see [Ref. 1, p. 213, Eqs. 6.22-6.25]). The velocity distribution function for each species is expanded in 8 *Gaussian Radial Basis Functions* (GRBFs) with density weights assigned to each. The GRBFs are essentially drifted Maxwellians at 8 "representative" 3D-velocity points of drift. The vector of 8 fluid moments has a pre-computed 8x8 analytic linear matrix $G_{8x8}$ relation to the vector of 8 GRBF density weighs at each point in 3D-real space. The time advanced 8 GRBF weights are computed from the explicitly time advanced moments using the inverse matrix $G_{8x8}^{-1}$. The components of the stress (and energy weighted stress) tensor "closure moments" are found from the 8 GRBF weights provided by analytically pre-computed matrices $P_{6x8}$ and $R_{6x8}$. The required velocity moments of the nonlinear Coulomb (inverse square) Fokker-Planck collision operator are evaluated from the GRBF weights using $O(10^3)$ velocity space grid integration over an large 8x8 pre-computed collision matrix $C_{8x8}$. Our choice of the GRBF representation was strongly motivated by the recent demonstration by Hirvijoki et al[6] that the Rosenbluth potential form[7] (see also [Ref. 1 p. 182] ) of the Landau collision operator can be analytically represented with GRBFs without numerical velocity derivatives.

Of course even a successful simulation with 8-moment and 8-GRBF velocity point system is unlikely to be physically convincing without a test of "numerical grid saturation". The 8 moment system is extended as needed to 12 (16, 20, ...) time evolved moment equations by adding energy $(1/2\ m\ v^2)^N$ factors to the energy density and 3 energy flux moments and two closure stress tensors with N=1 (2, 3, …) respectively. Clearly numerical practicality rests on an efficient choice for the corresponding number 8, or 12 (16, 20,…) of representative velocity 3D vector points or "grids". An appropriate normalization of the GRBF velocity grid is imperative otherwise the number of velocity grids and moments required would likely be impractical. Even in a small radial slice of a tokamak edge plasma, there is an enormous range of local particle velocities from the hot interior edge with low flows into the cold SOL with large (sound speed) flows at the divertor plate. The numerical simulations would start with a given "target" of toroidally symmetric local temperature and parallel field flow velocity 1D radial (or possibly 2D radial and poloidal) profiles for each species. The target profiles could come from experiment or be close to the expected final profiles. The 3D velocity grid "cloud" is to be locally normalized first shifting by the target local particle flow velocities and then normalizing to the target local thermal velocity. The 3D cloud of shifted and normalized velocity points are the same at each 3D spatial point and the same for each species with thermally normalized speeds ranging up to 1 or 2. The (up-front once) pre-computed "advance" matrices ($G_{8x8}$, $G_{8x8}^{-1}$, $P_{6x8}$, $R_{6x8}$, and $C_{8x8}$) are functions of the fixed target local temperature and velocity shift profiles. If the actual evolved and quasi-steady temperature and velocity flow profiles move too far from the target profiles, the advance matrices can easily be recomputed using the statistical quasi-steady profiles as new target profiles for normalization. Final results in a quasi-stationary state are expected to be independent of the details for the normalized velocity point cloud or the toroidally symmetric normalizing target profiles.

It is a common practice to locally normalize velocity space grids. For example in the global continuum delta-f gyrokinetic code GYRO [8, 9] this accounts for the unusually high efficiency of its global (full radius) simulations. Typically 8 energies normalized to the local species temperature is sufficient. Without local normalization, the number of velocity grids might need to be O(10-100) times more. The 8 energies combined with 8 pitch-angles (and 2 parallel direction) makes for a 128 "2D" gyrokinetic velocity space grid (or "particles per spatial cell"). The working hypothesis of the kinetic closed proposed here is that the invocation "fluid" moments will not require such a dense normalized velocity space grid. The 8-moment truncation is expected to saturate going to 12- (or 16-) moments with the corresponding number of normalized velocity grids.

The most difficult part of any numerical scheme for time advancing magnetized plasma equations is in how to find the electric field. Generalized cross magnetic field *drift* fluxes inversely proportional to the magnetic field are introduced. The generalized drift fluxes retain dependence on the full divergence of the stress tensors in addition to the electric and collisional forces. The time derivative of the generalized particle drift flux divided by the cyclotron frequency defines the polarization flux. As in the usual approach (see Ref. [3]) the *weak* drift approximation sets the exact cross field particle flux to the drift flux plus the polarization flux. The electric potential is then obtained from a time advanced generalized *vorticity* equation combining the particle drift flux and polarization density flux in the quasi-neutral charge continuity equation ($\vec{\nabla} \cdot \vec{j} = 0$). The computed ion polarization flux must be much less than the generalized drift flux to justify the *weak* drift approximation which follows only low drift frequency ion motion much less than the ion cyclotron. As in Ref [3] , following the explicit time advance of the vorticity equation, a *strong* drift approximation would set all cross field fluxes, including higher moment fluxes, equal to their corresponding *drift* fluxes without any explicit time advance. Because the electron cyclotron frequency is much higher than the ion cyclotron frequency and the electron polarization is negligible, the *strong* drift approximation is applied to the electrons from the beginning. From quasi-neutrality, the pure plasma ion density is set equal to the electron density obtained from the evolution of the electron particle continuity equation. After solution of the vorticity equation, the ion particle continuity equation is redundant. There is no gyro-averaging in this approach to properly cut-off short perpendicular motion at the ion gyroradius scale.

It may be possible to obtain a gyro-averaged perpendicular electric field by a novel method to advance the perpendicular electric field with an implicit time advance following the ion cyclotron motion with (and without) the *weak* drift approximation. The usual vorticity equation for the electric potential is by-passed. The parallel electric field (and hence the complete electric potential) is obtained from the quasi-neutral Poisson equation ($\vec{\nabla} \cdot \vec{E} = 0$). Extension to low-beta perpendicular magnetic perturbations with a parallel Ampere's law is straightforward.

The formulations for the GRBF representation, the moments hierarchy, the pre-computed time advance matrices relating GRBF weights to the time advance moments and the closure moments are given in Section II.A. The GRBF collisional moments are given in Sec. II.B. Appendix A provide some details on the conservation properties of the GRBF nonlinear Rosenbluth collision matrix. The generalized drift fluxes and evolution the electric potential via an extended vorticity equation are provided in Sec. II.C. A novel method to find the ion gyro-averaged potential is described in Sec. II.D. Extension to include magnetic field perturbations is given in Sec. II.E. Section III provides a discussion comparing and contrasting the GRBF kinetic closure presented here with the conventional Braginshii closure as formulated in Ref. [3]. The paper provides theory for a novel but testable numerical method. However readers looking for a demonstrated "proof of principle" will be disappointed. Numerical code tests of the GRBF kinetic closure are in progress.

## II. FORMULATION

### A. GRBF representation of the energy weighted time advanced moment hierarchy

The Vlasov equation for a plasma in an electric $\vec{E}$ and magnetic field $\vec{B}$ with collisions $C(f)$ and source $S$ is given by

$$\partial f / \partial t + \vec{\nabla} \cdot (\vec{v} f) + (Ze/m)(\vec{E} + \vec{v}/c \times \vec{B}) \cdot \vec{\nabla}_v f = C(f) + S \qquad [1]$$

$f(\vec{x}, \vec{v}, t)$ is the 6D kinetic distribution function. Ion and electron species labels are suppressed except where needed. The notation is standard or otherwise follows Ref. [1]. The distribution function is to be represented as a sum of weighted GRBF's or Maxwellians drifting at representative velocity points $v_i(\vec{x})$ [4]:

$$f(\vec{x}, \vec{v}, t) = \sum_i w_i(\vec{x}, t) f_i^{GRBF}(\vec{x}) = \sum_i w_i(\vec{x}, t)[(\gamma(\vec{x})/\pi)^{3/2} \exp[-\gamma(\vec{x})(\vec{v} - \vec{v}_i(\vec{x}))^2] \qquad [2]$$

where $\gamma = 1/v_{th}^2$ with $v_{th} = \sqrt{2T_0(\vec{x})/m}$ thermal velocity. The density weights $w_i(\vec{x}, t)$ have an index $i = [1, i_M]$ where $i_M$ is the number of independent moments time advanced. The velocity points are shifted by local velocity flows $\vec{u}_0(\vec{x})$ and normalized to the local thermal velocity: $\vec{v}_i(\vec{x}) = v_{th}(\vec{x})[\hat{\vec{v}}_i' + \hat{\vec{u}}_0(\vec{x})]$ to make them efficiently "representative". The local "target" temperature $T_0$ and flow velocity shifts $\vec{u}_0$ are considered constant or quasi-static in time and toroidally symmetric. Normally only the ion flow velocity along the toroidally symmetric equilibrium field $\vec{B}_0$ approaching the ion thermal speed is worth consideration. However, parallel electron sheath flow velocities approaching thermal speeds near the diverter plate may be of concern. The shifted and normalized velocity points $\hat{\vec{v}}_i'$ are independent of $(\vec{x}, t)$ and the same for each species. Going forward with the moment hierarchy, perturbations in the magnetic field are ignored: $\vec{B} \equiv \vec{B}_0$ and the electric field is electrostatic $\vec{E} = -\vec{\nabla}\Phi$. Parallel and perpendicular directions are given with respect to the local unperturbed equilibrium magnetic field $\vec{B}_0$. In Section E below, a simple extension to include the effects of perpendicular magnetic perturbations should suffice for low-β plasmas.

The first 4 moments for the particle density $n$ and particle fluxes $\vec{\Gamma} = n\vec{V}$ are given by the particle and momentum conservation equations

$$\partial n / \partial t + \vec{\nabla} \cdot \vec{\Gamma} = S_n \qquad [3]$$

$$m \partial \vec{\Gamma} / \partial t + \vec{\nabla} \cdot \vec{P} - Zen\vec{E} - Ze\vec{\Gamma} \times \vec{B}_0 / c = \vec{F}_C + \vec{S}_M \qquad [4]$$

where $\vec{F}_C$ is the exchange (friction) force and $\vec{P}$ is the symmetric stress tensor. It is useful to note $\vec{P} = p\vec{I} + \vec{\Pi} + mn\vec{V}\vec{V}$ where $p = nT$ is the pressure, and $mn\vec{V}\vec{V}$ is the dynamic stress. The second 4 energy moment conservation equations follow:

$$\partial U / \partial t + \vec{\nabla} \cdot \vec{Q} - Ze\vec{\Gamma} \cdot \vec{E} = \Delta_C + S_E \qquad [5]$$

$$\partial \vec{Q} / \partial t + \vec{\nabla} \cdot \vec{R} - (Ze/m)[U\vec{E} + \vec{P} \cdot \vec{E}] - (Ze/mc)\vec{Q} \times \vec{B}_0 = \vec{G}_C + \vec{S}_Q \qquad [6]$$

where $U = 3/2 p + mnV^2/2$ is the total energy density, $\Delta_C$ is the total energy exchange, and $\vec{Q}$ is the total energy flux. $\vec{R}$ is an energy weighted symmetric stress tensor, and $\vec{G}_C$ is an energy weighted friction. It is useful to note $\vec{Q} = \vec{q} + (5/2) p\vec{V} + \vec{\Pi} \cdot \vec{V} + (mnV^2/2)\vec{V}$ where $\vec{q}$ is the heat flux. The first 8 time advance moments $[n, \vec{\Gamma}, U, \vec{Q}]$ are linearly related to the 8 GRBF weights $w_i$ by

$$n = \int dv^3 f = \sum_i w_i \qquad [7a]$$

$$\vec{\Gamma} = \int dv^3 \vec{v} f = v_{th} \sum_i w_i \hat{\vec{v}}_i \qquad [7b]$$

$$U = \int dv^3 mv^2/2 f = (mv_{th}^2/2) \sum_i w_i [\hat{v}_i^2 + 3/2] \qquad [7c]$$

$$\vec{Q} = \int dv^3 (mv^2/2) \vec{v} f = (mv_{th}^2/2) v_{th} \sum_i w_i [\hat{v}_i^2 + 3/2] \hat{\vec{v}} \qquad [7d].$$

where $\hat{\vec{v}}_i = (\hat{\vec{v}}_i' + \hat{\vec{u}}_0)$. Inversion of the 8x8 matrix $G_{8x8}$ implied by Eq. [7] provides the GRBF weights from the time advance moments. Our experience has shown that to ensure the $[G_{8x8}]^{-1}$ inverse exists, the representative $\hat{\vec{v}}_i'$ should have differing speeds $|\hat{\vec{v}}_i'|$ to avoid any rotational symmetry which can make $G_{8x8}$ singular. Table 1. of Appendix A. gives an example of a specific choice for the case of 8 velocity points with the 3x8 velocity values chosen at random. As different choice would have a different spectrum of GRBF weights, but the physical simulations results for the moments should be insensitive to the detailed choice. Some choices may be better than others as to convergence on the number of GRBF weights (or moments). For example an even and wide spread over differing speeds (as in the example given) is likely superior.

The now time advanced weights are then used to the evaluate the linear closure moments:

$$\vec{\vec{P}} = \int dv^3 m\vec{v}\vec{v} f = mv_{th}^2 \sum_i w_i [\hat{\vec{v}}_i \hat{\vec{v}}_i + \vec{\vec{I}}/2] \qquad [8a]$$

$$\vec{\vec{R}} = \int dv^3 (mv^2/2)\vec{v}\vec{v} f = v_{th}^2 (mv_{th}^2/2) \sum_i w_i [(\hat{v}_i^2 + 7/2)\hat{\vec{v}}_i \hat{\vec{v}}_i + (\hat{v}_i^2/2 + 5/4)\vec{\vec{I}}] \qquad [8b]$$

The matrices $G_{8x8}$ and $[G_{8x8}]^{-1}$ implied by Eq. [7], and $P_{6x8}$ and $R_{6x8}$ implied by Eq. [8] are to be pre-computed. If the evolving quasi-steady toroidally symmetric temperature $T_0$ and flow velocity shifts $u_0$ profiles move too far from the starting "target" profiles, these matrices (as well as collision matrices $C_{8x8}$ below) will need to be re-computed (perhaps many times.)

Generalization to the next tranches of energy weighted time advance and closure moments is straightforward:

$$[U_N, \vec{Q}_N] = \int dv^3 [mv^2/2, (mv^2/2)\vec{v}](mv^2/2)^N f \qquad [9a]$$

$$[\vec{\vec{P}}_N, \vec{\vec{R}}_N] = \int dv^3 [m\vec{v}\vec{v}, (mv^2/2)\vec{v}\vec{v}](mv^2/2)^N f \qquad [9b]$$

The number of representative velocity points is increased to $i_M = 8+4N$ with an increasing number of velocity angles and speed selections available. The analytic $i_M \times i_M$ time advance matrices implied by analogy to Eq. [7] $G_{i_M \times i_M}$ and $[G_{i_M \times i_M}]^{-1}$ as well as the 2(N+1) 6 x $i_M$ matrices implied by analogy to Eq, [8] $P_{6xi_M}$ and $R_{6xi_M}$ are again pre-computed. Most importantly the form of any additional 4N time advance equations for $[U_N, Q_N]$ (i.e. like Eqs. [5,6] for $[U_0, \vec{Q}_0]$ ) with tensor closures $[\vec{\vec{P}}_N, \vec{\vec{R}}_N]$ (i.e. like $[\vec{\vec{P}}_0, \vec{\vec{R}}_0]$) remains unchanged: the number of terms is the same and the one stiff term $(Ze/mc)\vec{Q}_N \times \vec{B}_0$ driving fast ion cyclotron motion (like $(Ze/mc)\vec{\Gamma} \times \vec{B}_0$ in Eq. 4) remains in place. In principle this could allow a simple matrix inversion for an implicit advance of the stiff terms for all the flux $\vec{F}$ moments: $[\partial \vec{F}/\partial t - (Ze/mc)\vec{F} \times B_0]_{implicit} = [\ ]_{explicit}$ . If the 8-moment truncation fails to quickly saturate going to 12- ( or 16-) moments, it would seem that the proposed scheme is unlikely to be practical. For example, the GRBF representation of the next (N=1) tranche of 12 time advance moments is given by

$$U_1 = (m\upsilon_{th}^2/2)^2 \sum_i w_i [\hat{v}_i^4 + 3\hat{v}_i^2 + 3/4] \qquad [9c]$$

$$\vec{Q}_1 = (m\upsilon_{th}^2/2)^2 \upsilon_{th} \sum_i w_i [\hat{v}_i^4 + 4\hat{v}_i^2 + 5/2]\vec{\hat{v}}_i \qquad [9d]$$

and the closure moments by

$$\vec{\vec{P}}_1 = m\upsilon_{th}^2 (m\upsilon_{th}^2/2) \sum_i w_i [(\hat{v}_i^2 + 7/2)\vec{\hat{v}}_i\vec{\hat{v}}_i + (\hat{v}_i^2/2 + 5/4)\vec{\vec{I}}] \qquad [9e]$$

$$\vec{\vec{R}}_1 = \upsilon_{th}^2 (m\upsilon_{th}^2/2)^2 \sum_i w_i [(\hat{v}_i^4 + 9\hat{v}_i^2 + 63/4\vec{\hat{v}}_i\vec{\hat{v}}_i + (1/2\hat{v}_i^4 + 7/2\hat{v}_i^2 + 35/8)\vec{\vec{I}}] \qquad [9f]$$

Our first and now discarded approach to a GRFB kinetic closure focused on the general symmetric moment hierarchy (see Ref [1] p. 214). There are several reasons why the energy weighted moment hierarchy is clearly simpler and numerically more efficient than the general symmetric moment hierarchy: The first 13,40,121…symmetric moments have a decreasing 10, 20, 35…fraction of independent components yielding the corresponding number of independent representative velocity points. The symmetrization of the right hand sides leads to a rapidly increasing number of terms and multiplies. Going beyond the divergence of 2-tensor closure moments (in Eq. [4] and [6] with generalization in Eq. [9b]) to divergence of 3-,4-, 5-…tensor closures in the general hierarchy closure requires evaluating additional loops over Christoffel matrices at each stage.

**B. GRBF representation of the nonlinear Rosenbluth collisional moments**

Hirvijoki et al [6] recently demonstrated that the Rosenbluth potential form [7] (see also [Ref. 1 p. 182] ) of the Landau collision operator can be analytically represented with GRBFs without recourse to numerical velocity space derivatives. The equilibration of two widely separated 3D velocity space "balls" to a single Maxwellian "ball" $\partial f(\vec{v},t)/\partial t = C(f) \to C(f_{Max}) = 0$ was illustrated. Good number, momentum, and energy conservation was demonstrated with $O(10^3)$ "collocated" velocity points evaluting $f(\vec{v},t)$ represented by $O(10^3)$ GRBFs. The novelty here is that while integration over $O(10^3)$ "co-located" velocity point to get the moments of the collision operator ($\vec{F}_C$ in Eq. [4], $\Delta_C$ in Eq. [5], and $G_C$ (and generalizations $G_{CN}$) in Eq. [6]), only a few (maybe only 8 or 12) GRBF weights (and normalized 3D velocity points) are likely to be sufficiently accurate. Let $C_{ab}(\vec{v})$ correspond to the collision operator for species "a" on "b" with $\vec{v}$ the velocity space of species "a". In the GRBF representation

$$C_{ab}(\vec{v}) = \sum_{k,l} C_{ab}^{k,l}(\vec{v}) w_k^a w_l^b \qquad [10]$$

with (sums over like and unlike species as appropriate)

$$\vec{F}_{Ca} = \sum_{k,l} [\int d\upsilon^3 m\vec{\upsilon} C_{ab}^{k,l}(\vec{v})] w_k^a w_l^b = \sum_{k,l} \vec{F}_{Ca}^{k,l} w_k^a w_l^b = -\vec{F}_{Cb} \qquad [11a]$$

$$\Delta_{Ca} = \sum_{k,l} [\int d\upsilon^3 (m\upsilon^2/2) C_{ab}^{k,l}(\vec{v})] w_k^a w_l^b = \sum_{k,l} \Delta_{Ca}^{k,l} w_k^a w_l^b = -\Delta_{Cb} \qquad [11b]$$

$$\vec{G}_{CNa} = \sum_{k,l} [\int d\upsilon^3 \vec{\upsilon} (m\upsilon^2/2)^{(1+N)} (C_{ab}^{k,l}(\vec{v}) + C_{aa}^{k,l}(\vec{v}))] w_k^a w_l^b = \sum_{k,l} \vec{G}_{CNa}^{k,l} w_k^a w_l^b \qquad [11c]$$

The numerous and very expensive nonlinear collision matrices $[\vec{F}_{Ca}^{kl}(\vec{x}), \Delta_{Ca}^{kl}(\vec{x}), \vec{G}_{CNa}^{kl}(\vec{x})]$ are to be pre-computed. A detailed formulation of a test code for computation of $C_{ab}^{k,l}(\vec{v})$ in terms of Rosenbluth potentials is given in Appendix A. The accuracy of number conservation for 8 GRBFs is tested by defining an acceptable average error. Note there is of course no particle number collision moment in the density continuity equation Eq. [3], so that deviations from perfect conservation ( $\int d\upsilon_a^3 C_{ab}(\vec{\upsilon}_a) = 0$ independent of the GRBF weights) is not critical. This is an important advantage for the application of the GRBF representation to moments of the collision operator rather than directly to the kinetic distribution functions as in the purely kinetic approach exemplified by Ref. [6]. Non-conservation of particle number is precluded, and accuracy of the higher collisional moments is less important.

**C. The drift approximation and the evolution of the electrostatic potential**

The most naive path to find $\vec{E} = -\vec{\nabla}\Phi$ from the electrostatic potential $\Phi$ is via Poisson's equation $-\nabla^2 \Phi = 4\pi e(Zn_i - n_e)$ for local charge imbalance. However, this implies working on the very short Debye length scales not relevant to applications at hand. It is possible to work with an artificially much larger Debye length, then show the final results are insensitive to smaller lengths. A more conventional approach is to enforce the quasi-neutral approximation $Zn_i = n_e$ and then extract the $\Phi$ from the quasi-neutral charge or current conservation $\vec{\nabla} \cdot \vec{j} = \vec{\nabla} \cdot (Ze\vec{\Gamma}_i - e\vec{\Gamma}_e) = 0$. The most well traveled path in magnetized plasma physics is to cross the vector B-field with the flux equations to extract the cross-B or perpendicular drift fluxes. For example, crossing Eq. [4] with the B-vector defines a *generalized* drift flux $\vec{\Gamma}_{d\perp}$:

$$\vec{\Gamma}_{d\perp} \equiv nc\vec{b}_0 \times \vec{\nabla}_\perp \Phi / B_0 + c\vec{b}_0 \times [\vec{\nabla} \cdot \vec{P} - \vec{F}_{C\perp} - S_{M\perp}]/(ZeB_0) \qquad [12a]$$

$$\vec{\Gamma}_\perp \equiv \vec{\Gamma}_{d\perp} + 1/\omega_c \partial(\vec{b}_0 \times \vec{\Gamma}_\perp)/\partial t \qquad [12b]$$

where $\omega_c = (ZeB_0/mc)$ is the very high cyclotron frequency. There is no approximation in Eq [12b]. To avoid working on fast cyclotron time scales, the *weak* drift approximation substitutes $\vec{\Gamma}_{d\perp}$ for $\vec{\Gamma}_\perp$ in the time derivative term of the exact Eq. [12b]

$$\vec{\Gamma}_\perp \cong \vec{\Gamma}_{d\perp} + 1/\omega_c \partial(\vec{b}_0 \times \vec{\Gamma}_{d\perp})/\partial t \qquad [12c]$$

where $\vec{\Gamma}_{\perp pol} \equiv 1/\omega_c \partial(\vec{b}_0 \times \vec{\Gamma}_{d\perp})/\partial t$ is defined here as the polarization flux. As long as the time steps (*dt*) are large enough to avoid following the cyclotron motion, then $\vec{\Gamma}_{\perp pol} \ll \vec{\Gamma}_{\perp d}$. Substituting Eq. [12c] for the ions into the quasi-neutral charge or current conservation $\vec{\nabla} \cdot \vec{j} = 0$, dropping the small electron polarization so that $\Gamma_{\perp e} \cong \Gamma_{d\perp e}$, and also dropping electron viscosity and dynamic stress consistent with the small electron mass $m_e \ll m_i$ $\vec{\vec{P}}_e \cong p_e \vec{\vec{I}}$, we arrive at a generalized vorticity equation:

$$\partial \Omega_i / \partial t = \vec{\nabla} \cdot \vec{j}_\parallel + \vec{\nabla}_\perp \cdot c\vec{b}_0 \times [\vec{\nabla} \cdot (\vec{\vec{P}} + p_e\vec{\vec{I}})/B_0 - (S_{Mi} + S_{Me})/B_0] \qquad [13a]$$

where $\vec{j}_\parallel = \vec{b}_0 j_\parallel$ and the *extended* vorticity is defined by

$$\Omega_i \equiv \vec{\nabla}_\perp \cdot (1/\omega_{ci} B_0)[Zen\vec{\nabla}_\perp \Phi + \vec{\nabla}_\perp \cdot \vec{\vec{P}}_i - \vec{F}_{iC\perp} - S_{iM\perp}] \qquad [13b]$$

(Dropping small electron mass terms is not essential and could easily added back at little cost for the methods proposed here.) We refer to the *extended* vorticity because the "traditional" vorticity in Ref. [3] is defined by $\varpi \equiv \vec{\nabla}_\perp \cdot (1/\omega_{ci} B_0)[Zen\vec{\nabla}_\perp \Phi + \vec{\nabla}_\perp p_i]$. It would appear that *generalized* vorticity equation Eq [13] is actually equivalent to Eq. [76] of Ref [3]. Keeping the moment equations in the $\vec{\nabla} \cdot (\vec{V}n)$ conservative form (rather than $\vec{V} \cdot \vec{\nabla} n$ convective form) and avoiding any splitting of the stress tensor into pressure, viscous stress and dynamic stress, leads to the simpler form. Furthermore there is no need to discuss the "gyroviscous cancellation approximation" (see Ref. [3] p. 47530) or neglect any other terms.

There is an entirely analogous expression for the perpendicular energy drift flux (i.e. like Eqs [12]):

$$\vec{Q}_{d\perp} \equiv c\vec{b}_0 \times [(U\vec{E} + \vec{P} \cdot \vec{E}) - m(\vec{\nabla} \cdot \vec{R} - G_{C\perp} - S_{Q\perp})/Ze]/B_0 \qquad [14a]$$

and again without approximation from B cross on Eq. [6]

$$\vec{Q}_\perp = \vec{Q}_{d\perp} + 1/\omega_c \partial(\vec{b}_0 \times \vec{Q}_\perp)/\partial t \qquad [14b]$$

with a *weak* drift approximation analogous to Eq. [12c]

$$\vec{Q}_\perp \cong \vec{Q}_{d\perp} + 1/\omega_c \partial(\vec{b}_0 \times \vec{Q}_{d\perp})/\partial t \qquad [14c]$$

and similarly for the $\vec{Q}_{N\perp}$'s. Going forward after the generalized vorticity Eq. [13] is used to determine the electric field, the *strong* drift approximation sets all $[\vec{\Gamma}_\perp, \vec{Q}_{N\perp}]$ fluxes to their drift components $[\vec{\Gamma}_{d\perp}, \vec{Q}_{Nd\perp}]$ with no explicit time evolution for any perpendicular flux moment. While the strong drift approximation is natural for the electrons, it may be needed for the ions also to avoid any spuriously unstable cyclotron modes from using the GRBF kinetic closure. The parallel fluxes $[\vec{\Gamma}_\parallel, \vec{Q}_{N\parallel}]$ are advanced in time per Eqs. [4], [6], and analogs. The electron inertia term may be retained in the parallel momentum equation Eq.[4] with caution. The high frequency "electrostatic Alfven modes" $\omega/\omega_{ci} \sim (k_\parallel/k_\perp)\rho_* \sqrt{m_i/m_e}$ may become spuriously unstable. [Here $k_\parallel$ is a parallel and $k_\perp$ is a perpendicular wave number and $\rho_* = (c_s/\omega_{ci})/a$ is ion gyroradius relative to the minor radius.] This could be a problem for the *weak* drift approximation which requires frequencies less than the ion cyclotron frequency. Only the strong drift approximated electron continuity equation $\partial n_e/\partial t + \vec{\nabla}_\perp \cdot \vec{\Gamma}_{d\perp e} + \vec{\nabla} \cdot \vec{\Gamma}_\parallel = S_n$ is needed to advance the density for both species with quasi-neutraliy $Zn_i = n_e$. The ion continuity Eq. [3] is implicit in the vorticity Eq. [13] (equivalent to $\vec{\nabla} \cdot \vec{j} = 0$) but not explicitly used.

**D. Evolution of the electrostatic potential with gyro-averaging**

The most bothersome aspect of applying the *strong* drift approximation to the ions in particular is that gyro-averaging to properly treat and cut-off short wave motion on the ion gyro-radius ($\rho_i = \upsilon_{thi} / \omega_{ci}$) scale is precluded. (This is likely also the case using only the *weak* drift approximation consistently, i.e. Eqs. [12c], [14c], and higher moment flux analogs.) In the standard gyrokinetic approximation (see Ref. [1], p. 136), the gyro-averaging is done on the perpendicular space. Here we propose approximating the ion gyro-averaging by following the ion cyclotron motion. A new time step starts with a given electric field $\vec{E}$. The electrons are time-stepped with the *strong* drift approximation as above with no need to evolve the ion density continuity. In particular, the time advance of perpendicular part of the ion flux the follows the exact Eq. [4] with the given $\vec{E}$:

$$[\partial \vec{\Gamma}_{\perp i}(\vec{E}) / \partial t - \omega_{ci} \vec{\Gamma}_{\perp i}(\vec{E}) \times \vec{b}_0]_{implicit} = -\omega_{ci} \vec{\Gamma}_{d\perp i}(\vec{E}) \times \vec{b}_0]_{explicit} \quad [15a]$$

where $\vec{\Gamma}_{d\perp i}(\vec{E}_\perp)$ is given by Eq. [12b] (with $nc\vec{b}_0 \times \vec{\nabla}_\perp \Phi / B_0 = nc\vec{E}_\perp \times \vec{b}_0 / B_0$). The implicit time step in Eq. [15a] insures the ion cyclotron motion is being followed (at a fixed E-field) even though small time steps are not required. The remaining ion equations for $[U_i, \vec{\Gamma}_i, \vec{Q}_{iN}](\vec{E})$ are time stepped with fluxes in the implicit form using the same instantaneous starting $\vec{E}$. The next-step perpendicular electric field $\vec{E}'_\perp$ is obtained from implicitely time advancing $\vec{\Gamma}_{d\perp i}(\vec{E}'_\perp)$ using the *weak* drift approximation Eq. [12c] crossed with B and converted to implicit form:

$$[\partial \vec{\Gamma}_{d\perp i}(\vec{E}_\perp') / \partial t - \omega_{ci} \vec{\Gamma}_{d\perp i}(\vec{E}_\perp') \times \vec{b}_0]_{implicit} = -\omega_{ci} \vec{\Gamma}_{\perp i}(\vec{E}_\perp) \times \vec{b}_0]_{i explicit} \quad [15b]$$

with the time advanced $\vec{\Gamma}_\perp(\vec{E}_\perp)$ on the right-hand-side of Eq. [15b]. The left-hand-side $\partial \vec{\Gamma}_{di}(\vec{E}'_\perp) / \partial t$ must be smaller than the right-hand-side $\omega_{ci} \vec{\Gamma}_\perp(\vec{E}_\perp) \times \vec{b}_0$ to validate the *weak* drift approximation. Clearly Eq. [15a] and Eq. [15b] form a "leap-frog" system. Using B cross Eq. [12a], the next-step perpendicular electric field is

$$\vec{E}'_\perp = \vec{\Gamma}_{\perp d}(\vec{E}'_\perp)[B_0 / nc] + [\vec{\nabla} \cdot \vec{P}(\vec{E}) - \vec{F}_{c\perp}(\vec{E}) - S_{M\perp}] / (nZe) \quad [15c]$$

The next-step parallel electric field $\vec{E}'_\parallel = -\nabla_\parallel \Phi'$ is obtained by inverting the quasi-neutral Poisson equation $\vec{\nabla} \cdot \vec{E} = 0$:

$$\vec{\nabla} \cdot (\vec{b}_0 E'_\parallel) = -\vec{\nabla}_\perp \cdot \vec{E}'_\perp \quad [15d]$$

In this novel method, the generalized vorticity equation Eqs. [13] for advancing the electric field has been replaced by the exact equations Eqs. [4,5,6,9] responding to the same instantaneous electric field with the ion cyclotron motion included.

Again only the electron density continuity equation is used with $\partial n_e / \partial t + \vec{\nabla}_\perp \cdot \vec{\Gamma}_{d\perp e} + \nabla \cdot \vec{\Gamma}_{\parallel e} = S_n$. Substituting into the unused density continuity Eq. [3] for the ions and subtracting the quasi-neutral Eq. [3] for the electrons, the quasi-neutral charge continuity is $\vec{\nabla} \cdot \vec{j} = \nabla \cdot [Ze\vec{\Gamma}_i - e(\vec{\Gamma}_{\perp de} + \vec{\Gamma}_{\parallel e})]$. There is a paradox: Since $\vec{\nabla} \cdot \vec{E} = 0$ Eq. [15d] was used rather than $\vec{\nabla} \cdot \vec{j} = 0$, the latter is unlikely to hold with any accuracy. However the same paradox arises in the more conventional way of finding the electric field from the vorticity equation $\vec{\nabla} \cdot \vec{j} = 0$ Eq. [13]: the quasi-neutral $\vec{\nabla} \cdot \vec{E} = 0$ is unlikely to hold with any accuracy.

**E. Extensions to include magnetic perturbations for a low-β plasma**

For a low beta plasma, typical of the edge and scape-off-layer of a tokamak plasma, a straightforward and consistent way to include small perpendicular magnetic field perturbations $|\delta\vec{B}_\perp| << B_0$ while deprecating any importance of parallel perturbations $\delta B_\parallel$ follows from the *ansatz*

$$\delta\vec{B} = \vec{\nabla} \times (\vec{b}_0 \delta A_\parallel) \qquad [16a]$$

neglecting $\delta\vec{A}_\perp$ entirely. This form of the low-β approximation satisfies $\vec{\nabla} \cdot \delta\vec{B} = 0$ (with the MHD equilibrium $\vec{\nabla} \cdot \vec{B}_0 = 0$ of course). Again, parallel and perpendicular directions refer to the unperturbed magnetic field $\vec{b}_0$ direction. The parallel electric field is $E_\parallel = -\nabla_\parallel \Phi - 1/c \, \partial \delta A_\parallel / \partial t$ and the perpendicular field remains electrostatic with $\vec{E}_\perp = -\vec{\nabla}_\perp \Phi$ (consistent with neglect of $\delta\vec{A}_\perp$). Using the evolving internal parallel current density moment $j_\parallel$, Ampere's law to obtain $\delta A_\parallel$ is then written

$$\nabla^2 \delta A_\parallel - \vec{b}_0 \cdot \vec{\nabla}(\vec{\nabla} \cdot \vec{b}_0 \delta A_\parallel) = -4\pi(j_\parallel - \vec{j}_0 \cdot \vec{b}_0) \qquad [16b]$$

Eq. [16b] is consistent with Eq. [28] of Ref. [3], if the left-hand-side is interpreted as the the "fast" derivative $\nabla_f^2 \delta A_\parallel$. $\vec{j}_0 = \nabla \times \vec{B}_0 / (4\pi/c)$ is the equilibrium current. It follows exactly from Eq. [16a] that

$$\delta \vec{B}_\perp = (\vec{\nabla}\delta A_\parallel) \times \vec{b}_0 + \delta A_\parallel [(\vec{\nabla} \times \vec{b}_0) - \vec{b}_0(\vec{b}_0 \cdot \vec{\nabla} \times b_0)] \qquad [16c]$$

Eq. [16b] is consistent with Eq. [26] of Ref. [3] if the right-hand-side is the defined use of the "fast" derivative $(\vec{\nabla}_f \delta A_\parallel) \times \vec{b}_0$. It also follows exactly that

$$\delta B_\parallel = \delta A_\parallel (\vec{b}_0 \cdot \vec{\nabla} \times \vec{b}_0) \qquad [16d]$$

Where it should be clear that $\delta B_\parallel <<|\vec{B}_\perp|$.

The perturbed field can be simply added to the flux equations with $\vec{B}_0$ replaced by $\vec{B}_0 + \delta\vec{B}$ in Eqs. [4,6]. By the same "cross $\vec{B}_0$" steps, the generalized drift fluxes Eqs. [12a,14a] acquire "magnetic flutter" additions:

$$\vec{\Gamma}_{d\perp} \Rightarrow \vec{\Gamma}_{d\perp}^{(12a)} + \Gamma_\parallel \delta\vec{B}_\perp / B_0 - \vec{\Gamma}_\perp \delta B_\parallel / B_0 \approx \vec{\Gamma}_{d\perp}^{(12a)} + \Gamma_\parallel \delta\vec{B}_\perp / B_0 \qquad [17a]$$

$$\vec{Q}_{d\perp} \Rightarrow \vec{Q}_{d\perp}^{(14a)} + Q_\parallel \delta\vec{B}_\perp / B_0 - \vec{Q}_\perp \delta B_\parallel / B_0 \approx \vec{Q}_{d\perp}^{(14a)} + Q_\parallel \delta\vec{B}_\perp / B_0 \qquad [17b]$$

where $\delta B_\parallel$ terms can be safely neglected, e.g. $\vec{\Gamma}_\perp \delta B_\parallel / B_0 \sim \vec{\Gamma}_{d\perp} \delta B_\parallel / B_0 << \Gamma_\parallel \delta\vec{B}_\perp / B_0$. Using the *weak* drift approximation Eq. [12c], the generalized vorticity equation Eq. [13a] in Sec. C now includes the magnetic flutter current

$$\partial \Omega_i / \partial t = \vec{\nabla} \cdot \vec{j}_\parallel + \vec{\nabla}_\perp \cdot (j_\parallel \delta\vec{B}_\perp / B_0) + \vec{\nabla}_\perp \cdot c\vec{b}_0 \times [\vec{\nabla} \cdot (\vec{P} + p_e \vec{I}) / B_0 - (S_{Mi} + S_{Me})/B_0] \qquad [18a]$$

with the generalized vorticity including the magnetic field perturbation:

$$\Omega_i = \vec{\nabla}_\perp \cdot (1/\omega_{ci} B_0)\{Zen[\vec{\nabla}_\perp \Phi + (\Gamma_\parallel^i / nc)\vec{b}_0 \times \delta\vec{B}_\perp] + \vec{\nabla}_\perp \cdot \vec{P} - \vec{F}_{iC\perp} - S_{iM\perp}\} \qquad [18b]$$

where $\vec{\Gamma}_\perp \delta B_\parallel / B_0 << \Gamma_\parallel \delta\vec{B}_\perp / B_0$ has been dropped. Note that the combination $\vec{b}_0 \times [\vec{\nabla}_\perp \Phi + (\Gamma_\parallel^i / nc)\vec{b}_0 \times \delta\vec{B}_\perp] \approx \vec{b}_0 \times [\vec{\nabla}_\perp \Phi - (\Gamma_\parallel^i / nc)\vec{\nabla}_{f\perp} \delta A_\parallel]$, which also appears in the extended $\vec{\Gamma}_{d\perp}$ drift flux Eq. [17a], is reminiscent of the generalized potential $\delta U$ often used in $\delta f$–gyrokinetic codes for microturbulence where perpendicular derivatives are "fast" and parallel derivatives are "slow": $\vec{b}_0 \times \vec{\nabla}_{f\perp} \delta U = \vec{b}_0 \times \vec{\nabla}_{f\perp}[\delta\Phi - (v_\parallel / c)\delta A_\parallel]$ where the parallel particle velocity has been replaced by a fluid velocity $\Gamma_\parallel / n$. The *strong* drift approximation again sets all $[\vec{\Gamma}_{N\perp}, \vec{Q}_{N\perp}]$ to $[\vec{\Gamma}_{d\perp}, \vec{Q}_{Nd\perp}]$ with no explicit time evolution for any perpendicular flux moment. $Ze(\vec{b}_0 \cdot \delta\vec{\Gamma}_{d\perp}^{(12a)} \times \delta\vec{B}_\perp)$ and $(Ze/mc)(\vec{b}_0 \cdot \delta Q_{d\perp}^{(14a)} \times \delta\vec{B}_\perp)$ should be added to the right-hand-side of the explicitly time advanced parallel momentum and energy flux Eqs. [4] and [6[ respectively.

## III. SUMMARY AND DISCUSSION

The novel GRBF kinetic fluid moment closure scheme presented here is in marked contrast to the commonly used Braginskii [2] collisional two fluid closure commonly used in tokamak edge and scrape-off-layer plasma codes. The detailed closed form of the Braginskii closure equations with the *strong* drift approximation is detailed in Ref. [3]. In the primary GRBF kinetic closure case, 8 moments $[n,\vec{\Gamma},U,\vec{Q}]$ for each species are time advanced (see Eq. [7]) in conservative form [e.g. $\vec{\nabla}\cdot(n\vec{V})$, $\vec{\nabla}\cdot\vec{\vec{P}}$] in place of 5 moments (equivalent to $[n,\vec{\Gamma},U]$) in convective form [e.g. $\vec{V}\cdot\vec{\nabla}n$, $m\vec{V}\cdot\vec{\nabla}\vec{V}$]. In contrast to the Braginskii closure, neither the primary kinetic closure moments for stress and energy weighted stress tensor $[\vec{\vec{P}},\vec{\vec{R}}]$ (see Eq. [8]) nor the collisional source moments $[\vec{F}_C,\Delta_C,\vec{G}_C]$ (see Eq. [11]) depend *explicitly* on gradients of time advance lower moments (like $[n,\vec{\Gamma},U]$). For example in the Braginskii closure, the most troublesome closure for the viscous stress tensor $\vec{\vec{\Pi}} = -\vec{\vec{\eta}}:\vec{\nabla}\vec{V}+..$ (e.g. see Eq. [14] of Ref. [3]) is broken free from the total stress tenor and depends on the heat conduction part of the Braginskii closure for $\vec{Q}$: $q_{\parallel} = -\kappa_{\parallel}\nabla_{\parallel}T+..$ etc. Most importantly, the GRBF kinetic closure is not limited to high collisionality.

As in the conventional approach (like Ref. [3]), the electric field is found from a quasi-neutral current continuity equation $\vec{\nabla}\cdot\vec{j}=0$ with a g*eneralized* vorticity equation (see Eq. [13]). The *weak* drift approximation is used where the cross field ion flux is broken into a drift and polarization fluxes $\vec{\Gamma}_{\perp} \approx \vec{\Gamma}_{d\perp} + \vec{\Gamma}_{\perp pol}$ with $\vec{\Gamma}_{\perp pol} = 1/\omega_c \partial(\vec{b}_0 \times \vec{\Gamma}_{d\perp})/\partial t$ Eq. [12c] and $\Gamma_{d\perp}$ is a *generalized* drift (see Eq. [12a]) with $\vec{\nabla}\cdot\vec{\vec{P}}$ replacing $\vec{\nabla}\cdot(p\vec{\vec{I}})$ in the usual diamagnetic drifts. This avoids difficult to verify approximations involving $\vec{\nabla}\cdot\vec{\vec{\Pi}}$ and $\vec{\nabla}\cdot(mn\vec{V}\vec{V})$ in Ref. [3] and complicated general expression of the gyroviscous force in Ref. [10]. Compare the simple looking vorticity Eq. [13] with equation Eq. [76] and Apprendix D derivation of Ref. [3]. The *strong* drift approximation $\vec{\Gamma}_{\perp} \approx \vec{\Gamma}_{d\perp}$ and $\vec{Q}_{\perp} \approx \vec{Q}_{d\perp}$ (Eq. [14]) is applied to the electrons with the electron continuity Eq. [3] evolving the quasi neutral density $n = n_e = n_i$. The vorticity equation Eq. [13] subsumes the ion continuity equation. Some preliminary test code results with linearized adiabatic electrons and GRBF ion equations, suggest spurious high frequency ion cyclotron modes may obtain. Going beyond the vorticity equation which rests on *weak* drift approximation, it may be necessary to apply the *strong* drift approximation to eliminate such spurious modes. As in Ref. [3], the ions are then purely "drift kinetic" which has no ion gyro-radius cut-off at short perpendicular wave lengths. Only the parallel flux moment equations $[\vec{\Gamma}_\|, \vec{Q}_\|]$ Eqs. [4] (including electron inertia) and [6], as well as the (electron) density $n_e$ and total energy moments U Eqs. [3] and [4], are explicitly time evolved (including electron inertia).

To retain the ion gyro-radius cut-off and gyro-averaging of the electric potential, the $\vec{\nabla} \cdot \vec{j} = 0$ and vorticity equation (as commonly used, e.g. Ref. [3]) can be abandoned in favor of a gyro-averaged electric field obtained from the quasi-neutral Poisson equation $\vec{\nabla} \cdot \vec{E} = 0$. A novel two-half-step "leap-frog" method (see Eq. [15]) was proposed which avoids the *strong* drift approximation for the ions: gyro-averaging obtains from following the ion-cyclotron motion implicitly in evolving the perpendicular flux $[\vec{\Gamma}_\perp, \vec{Q}_\perp]$ equations Eqs. [4] and [6].

Simple extensions of Sections II.A and II.C to include the magnetic field perturbations for a low-β plasma were provided in Sec. II. E. $\vec{B}_0$ is replaced by $\vec{B}_0 + \delta\vec{B}$ in the perpendicular flux equations Eqs. [4] and [6]. The derivation follows by referring parallel and perpendicular field directions to the unperturbed field direction $\vec{b}_0$. "Magnetic flutter" fluxes $[\Gamma_\| \delta\vec{B}_\perp / B_0, Q_\| \delta\vec{B}_\perp / B_0]$ are naturally added to the extended drift fluxes $[\vec{\Gamma}_{d\perp}, \vec{Q}_{d\perp}]$ (see Eq. [17]). A simple *ansatz* approximation that $\delta\vec{B} = \vec{\nabla} \times (\vec{b}_0 \delta A_\|)$ with $\delta A_\perp = 0$ [Eq. 16a] follows Faraday's law for the electric field and allows for a rigorously defined Ampere's law for $\delta A_\|$ (see Eq. [16b]). For example, the "fast" derivatives used in the Ref. [3] Ampere's law are now clearly defined: $\nabla_f^2 \delta A_\| = \nabla^2 \delta A_\| - \vec{b}_0 \cdot \vec{\nabla}(\vec{\nabla} \cdot \vec{b}_0 \delta A_\|)$. The effects of $\delta B_\|$ can be deprecated in low-β plasmas.

**Acknowledgment**

This work was supported by the U.S. Department of Energy. Office of Fusion Energy Sciences, Theory Program grant award DE-FG02-95ER54309. One of us (ZD) acknowledges a General Atomics sponsored ORISE postdoctoral fellowship.

**Disclaimer**

**Appendix A: Formulation of the nonlinear Rosenbluth GRBF collision operator**

The Landau Rosenbluth, MacDonald [7] and Judd Coulomb collision operator for species "a" with velocity $\vec{v}$ colliding on "b" can be written as (see Ref. [1] p183)

$$C_{ab} = [(\gamma_{ab}/2)/m_a][\partial^2/\partial v_\alpha \partial v_\beta (f_a \partial^2 G_b/\partial v_\alpha \partial v_\beta)] - 2(1+m_a/m_b)\partial/\partial v_\alpha (f_a \partial H_b/\partial v_\alpha)] \quad [A.1]$$

where the famous Rosenbluth potenials are defined by $G_s(\vec{v}) = \int dv'^3 \vec{v}' f_s(\vec{v}')u$ and $H_s(\vec{v}) = \int dv'^3 \vec{v}' f_s(\vec{v}')/u$ where $u = |\vec{v} - \vec{v}'|$. Using the identities $\partial^2 G_s/\partial v_\alpha \partial v_\alpha = 2H_s$ and $\partial^2 H_s/\partial v_\alpha \partial v_\alpha = -4\pi f_s$, Eq. [A.1] can be written[6] in a form with fewer high derivatives as

$$C_{ab} = L_{ab}[(m_a/m_b)f_a f_b + \mu_{ab}(\partial f_a/\partial v_\alpha)(\partial \varphi_b/\partial v_\alpha) - (\partial^2 f_a/\partial v_\alpha \partial v_\beta)(\partial^2 \psi_b/\partial v_\alpha \partial v_\beta)] \quad [A.2]$$

where $\varphi_b = -H_b/4\pi = (n_b/v_{thb})\Phi(s_b)$, $\psi_b = -G_b/8\pi = (n_b v_{thb})\Psi(s_b)$, $\mu_{ab} = m_a/m_b - 1$, and $s_b = v/v_{thb} = \sqrt{\gamma_b}v$. In Gaussian units $L_{ab} = 4\pi\gamma_{ab}/m_a = (4\pi)^2(e_a^2 e_b^2/m_a^2)\ln\Lambda$ with $\ln\Lambda$ the Coulomb logarithm. Most importantly there are well know analytic forms for the Rosenbluth potentials: $\Phi(s) = erf(s)/s$ and $\Psi(s) = [s+1/(2s)]erf(s) + \exp(-s^2)/2\pi$. G corresponds to velocity space diffusion and H corresponds to drag.

A code has been written to test the particle (density) conservation properties of the GRBF collision moments matrix $C_{ab}^{k,l}(\vec{v}_a)$ defined in Eq. [10]: $C_{ab}(\vec{v}) = \sum_{k,l} C_{ab}^{k,l}(\vec{v})w_k^a w_l^b$. The details of the test are formulated here. Following Eq. [1] and expanding Eq. [A.1], we have

$$C_{ab}^{k,l}(\vec{v}_j) = L_{qb}[(m_a/m_b)F_a^k F_b^l + \mu_{ab}(\partial F_a^k/\partial v_\alpha)(\partial \varphi_b^l/\partial v_\alpha) - (\partial^2 F_a^k/\partial v_\alpha \partial v_\beta)(\partial^2 \psi_b^l/\partial v_\alpha \partial v_\beta)] \quad [A.3]$$

where

$$\begin{aligned}
&F_a^k(s_a^{j,k}) = (\gamma_a/\pi)^{3/2} \exp[-(s_a^{j,k})^2] \\
&\partial F_a^k/\partial v_\alpha = (s_a^{j,k}/v_{tha})(\hat{\vec{v}}_a^j - \hat{\vec{v}}_a^k)_\alpha [\partial F_a^k(s_a^{j,k})/\partial s_a^{j,k}] \\
&\partial^2 F_a^k/\partial v_\alpha \partial v_\beta = \gamma_a \delta_{\alpha\beta}[(1/s_a^{j,k})\partial F_a^k(s_a^{j,k})/\partial s_b^{j,l}] + \\
&\quad \gamma_a[\hat{\vec{v}}_a^j - \hat{\vec{v}}_a^k]_\alpha [\hat{\vec{v}}_a^j - \hat{\vec{v}}_a^k]_\beta (1/s_a^{j,k})\partial[(1/s_a^{j,k})\partial F_a^k(s_a^{j,k})/\partial s_a^{j,k}]/\partial s_a^{j,k}
\end{aligned} \quad [A.4]$$

and where $s_a^{j,k} = |\vec{\hat{v}}_a^j - \vec{\hat{v}}_a^k|$. As in the text $\vec{\hat{v}}_a^k = \vec{\hat{v}}'^k + \vec{\hat{u}}_{0a}$ with k=1,i$_{max}$ with the velocities normalized to $v_{tha}$. However $\vec{\hat{v}}_a^j = \vec{\hat{v}}''^j + \vec{\hat{u}}_{0a}$ corresponds to a much larger collisional moment integration grid $\vec{\hat{v}}''^j$ with j=1,j$_{max}$ and centered on the normalized colliding species "a" flow velocity $\hat{u}_{0a}$. Similarly

$$F_b^l(s_b^{j,l}) = (\gamma_b / \pi)^{3/2} \exp[-(s_b^{j,l})^2]$$
$$\partial \varphi_b^l / \partial v_\alpha = \gamma_b [\vec{\hat{v}}_a^j(v_{tha}/v_{thb}) - \vec{\hat{v}}_b^l][\partial \Phi(s_b^{j,l})/\partial s_b^{j,l}]$$
$$\partial^2 \psi_b^l / \partial v_\alpha \partial v_\beta = \gamma_b v_{thb} \delta_{\alpha\beta}[(1/s_b^{j,l})\partial \Psi(s_b^{j,l})/\partial s_b^{j,l}] + \quad [A.5]$$
$$\gamma_b v_{thb}[\vec{\hat{v}}_a^j(v_{tha}/v_{thb}) - \vec{\hat{v}}_b^l]_\alpha[\vec{\hat{v}}_a^j(v_{tha}/v_{thb}) - \vec{\hat{v}}_b^l]_\beta (1/s_b^{j,l})\partial[(1/s_b^{j,l})\partial \Psi(s_b^{j,l})/\partial s_b^{j,l}]/\partial s_b^{j,l}$$

with $s_b^{j,l} = |\vec{\hat{v}}_a^j(v_{tha}/v_{thb}) - \vec{\hat{v}}_b^l|$ and as in the text $\vec{\hat{v}}_b^l = \vec{\hat{v}}'^l + \hat{u}_{0b}$. Note that unlike $s_a^{j,k}$, the flow velocities in $s_b^{j,l}$: $\vec{u}_{0a}/v_{thb} - \vec{u}_{0b}/v_{thb}$ is not zero in general.

The conservation of particles corresponds to $\int dv_a^3 C_{ab}(\vec{v}_a) = 0$. The numerical equivalent $\sum_{k,l} w_k^a w_l^b [\sum_j (\Delta v_a)^3 \sum_{k,l} C_{ab}^{k,l}(\vec{v}_a^j)]$ can not be perfectly zero for an arbitrary set of weights $w_k^a w_l^b$. That would require $[\sum_j (\Delta v_a)^3 \sum_{k,l} C_{ab}^{k,l}(\vec{v}_a^j)] \equiv 0$ for all k and l. To define an acceptable error, the integral over the first term in Eq. [A.3] C(1)=$[\sum_j (\Delta v_a)^3 \sum_{k,l} {}^{FF}C_{ab}^{k,l}(\vec{v}_a^j)]$ which is guaranteed positive must sufficiently balance out the integral of the second and third term C(2,3)= $[\sum_j (\Delta v_a)^3 \sum_{k,l} ({}^{F\varphi}C_{ab}^{k,l}(\vec{v}_a^j) + {}^{F\psi}C_{ab}^{k,l}(\vec{v}_a^j))]$ as a fraction of the first. $E(k,l) = [C(1) + C(2,3)]/C(1)$ is the error in one k,l term. The test code considers the k and l grid to i$_{max}$ = 8 moment or 8 GRBF's with the 8 representative velocities $\vec{\hat{v}}'^k$ having speeds about equally spaced up to about 2. This resulted from 3x8 velocity values chosen by random. Results here are rather insensitive to how the representative velocity points are selected. (See Table 1.) More importantly practicality requires that the integration velocity space be kept reasonably small. A 3D equally spaced cube of grids truncated to a sphere of $\vec{\hat{v}}''^j$ moment integration points with maximum speeds up to about 6 and total j$_{max}$ of O(1500) was deemed sufficient. Note again that very expensive nonlinear collision matrices $[\vec{F}_{Ca}^{kl}(\vec{x}), \Delta_{Ca}^{kl}(\vec{x}), \vec{G}_{CNa}^{kl}(\vec{x})]$ in Eq [11] are to be pre-computed and don't need re-computing with the evolving moment time steps. This should be contrasted with the Ref. [4] purely kinetic

GRBF test case which found that [O(1500)]-*cubed* velocity points needed to be followed in time for almost perfect conservation of number, momentm and energy. In the GRBF kinetic moments closure here, only 8 (12, 16 etc) velocity points are followed in time and the O(1500) velocity grid integrations are done only once. Because there is no actual collisional moment in the particle continuity equation Eq. [3], perfect conservation in momentum and energy is less important.

First consider like species collisions where $\mu_{ab} = 0$ and the second "$F\psi$" term vanishes. The test code demonstrated $C_{aa}^{kk}(\vec{v}_a^j) = 0$ (within round-off) for all k=l and all j as required: A single shifted Maxwellian annihilates the collision operator. It then follows that on integration the $E(k,k)$ error is zero within round-off. For the general $E(k,l)$ the average $\pm$ deviations from zero is less than 3% i.e. $\sum_{k,l} E(k,l) / \sum_{k,l} 1 \approx 2.7\%$. Of course there no is actual (like-like or like-unlike) collision term in the particle density equation or like-like collision for momentum or energy moments, i.e. non-need be computed. The higher collision moments $\vec{v}(mv^2/2)^{1+N}$ for the energy weighted friction $\vec{G}$ [11c] (and so on $(mv^2/2)(mv^2/2)^{1+N}$) with N=0,1,2.. are the only places conservation errors in like-like collisions can have any effect. Of course is possible to correct any collision operator for number, momentum and energy conservation. The momentum and energy exchanges Eqs. [11a,11b] depend on collision of unlike species to which we now turn.

For collisions with unlike species as in the "exchange terms", it is important to have the massive and slow species (i.e. ions) to be the colliding thene species "a" so that $m_a/m_b$ is huge (like $60^2$) and $v_{tha}/v_{thb}$ is very small (like 1/60). When evaluating $C_{ie}$, the "FF" term then nearly cancels the "$F\phi$" "drag" term and the "$F\psi$" "diffusion" term is small. The average deviation from zero $\sum_{k,l} E(k,l) / \sum_{k,l} 1 \approx 0.15\%$ is very small for ions colliding on electrons. In contrast, if the fast species is the colliding species "a" then $v_{tha}/v_{thb}$ is very large and hence $s_b^{j,l} = |\vec{\hat{v}}_a^j (v_{tha}/v_{thb}) - \vec{\hat{v}}_b^l|$ is very large. This makes $F_b^l(s_b^{j,l}) = (\gamma_b/\pi)^{3/2} \exp[-(s_b^{j,l})^2]$ vanishingly small and impossible to distinguish from zero even with double or quadruple precision. While it may seem strange to think of ions colliding on electrons, this is really not a problem for the momentum and energy exchange terms Eqs. [11a,11b], since $F_{Ce} = -F_{Ci}$ and $\Delta_{Ce} = -\Delta_{Ci}$. The problem of small colliding mass first appear in evaluating $\vec{G}_{Ce}$ Eq. [11c] (and higher moments) which require $C_{ei}$ (in addition to $C_{ee}$ for which there is no small mass ratio problem).

It would appear that use of the small mass ratio approximation (Eq. 5.57 p185 of Ref. [1]):

$$C_{ei} \approx (\gamma_{ei} n_i / 2m_e) \partial / \partial v_\alpha [U^i_{\alpha\beta} \partial f_e / \partial v_\beta] \qquad [A.6]$$

where $U^i_{\alpha\beta} = 1/u_i^3 [u_i^2 \delta_{\alpha\beta} - u_{i\alpha} u_{i\beta}]$ with $\vec{u}_i = \vec{v} - \vec{V}_i$ is unavoidable. The approximated ion species should be close to a drifted Maxwellian. It known for example that the small mass approximation is not accurate for the energy exchange $\Delta_{Ce} = -\Delta_{Ci}$ (Ref. [1] p. 186). The GRBF $C_{ei}$ form is now linear in the electron weights $C_{ei}(\vec{v}_e^j) \approx \sum_{k,l} C_{ei}^k(\vec{v}_e^j) w_k^e$ :

$$C_{ei}^k(\vec{v}_e^j) = (\gamma_{ei} n_i / 2m_e)[(\partial U^i_{\alpha\beta} / \partial v_\beta)(\partial F_e^k / \partial v_\alpha) + U^i_{\alpha\beta}(\partial^2 F_e^k / \partial v_\alpha \partial v_\beta)] \qquad [A.7a]$$

where (with "a"="e") $\partial F_e^k / \partial v_\alpha$ and $\partial^2 F_e^k / \partial v_\alpha \partial v_\beta$ are given by Eq. [A.4]. $\vec{v}_e^j = \vec{v}^{n,j} + \vec{u}_{0e}$ and

$$U^i_{\alpha\beta} = (1/v_{the})\{\hat{V}^j_{\alpha\beta} + [\hat{v}^j_{e\alpha} \hat{V}_{i\beta} + \hat{v}^j_{e\beta} \hat{V}_{i\alpha} + (\delta_{\alpha\beta} - 3(\hat{v}^j_{e\alpha}/\hat{v}^j_e)(\hat{v}^j_{e\beta}/\hat{v}^j_e))(\vec{\hat{v}}^j_e \cdot \vec{V}_i)](v_{thi}/v_{the})/(\hat{v}^j_e)^2 \qquad [A.7b]$$

where $\hat{V}^j_{\alpha\beta} = [(\hat{v}^j_e)^2 \delta_{\alpha\beta} - \hat{v}^j_{e\alpha} \hat{v}^j_{e\beta})]/(\hat{v}^j_e)^3$. After some tedious algebra, it can be shown that

$$\partial U^i_{\alpha\beta} / \partial v_\beta \equiv 0 \qquad [A.7c]$$

Note that the ion moments with the corresponding GRBF small mass ratio approximation $C_{ie}^k(\vec{v}_i^j) = -(\gamma_{ie} n_i / 2m_e) U^e_{\alpha\beta}(\partial^2 F_i^k / \partial v_\alpha \partial v_\beta)$ analogous to Eq. [A.7a] (with $U^e_{\alpha\beta}$ analogous to $U^i_{\alpha\beta}$ interchanging "i" for "e" in Eq. A.7b]), might be usefully compared those using the full GRBF given by Eqs. [A.3, A.4, A.5] with "a"="i" and "b=e". This is particularly the case for the friction force $F_{Ci}$ where the small mass ratio approximation may be adequate.

**Table 1.** Example 8x3 GRBF normalized velocity grids chosen by random in each direction then corrected so the average velocity in each direction is null. i=1,8

| $\hat{v}_x(i)$ | $\hat{v}_y(i)$ | $\hat{v}_z(i)$ |
|---|---|---|
| -1.53635037124146 | 0.415647106806283 | -0.211854701108601 |
| -1.45991021922900 | 0.526135507783161 | -0.963228984386408 |
| -0.478803063718722 | -0.936815843803036 | 1.61218597724140 |
| 0.464391897070831 | 0.641607873799124 | -0.819406362261873 |
| 1.35281504818204 | -1.70218885308646 | -0.991442011970825 |
| 0.978513062896024 | 0.692907595900885 | -0.857307770869768 |
| -0.530286416562434 | 0.130992784169217 | 0.564080765271496 |
| 1.20963006260272 | 0.231713828430825 | 1.66697308808458 |

| $|\hat{v}(i)|$ |
|---|
| 1.60562056382946 |
| 1.82646283759664 |
| 1.92510257499195 |
| 1.13962594051790 |
| 2.38966790773537 |
| 1.47396925500805 |
| 0.785206917216288 |
| 2.07260595942320 |